\documentstyle{article}
\begin{document}
Critical Coupling in (1+1)-Dimensional Light-Front $\phi^{4}$ Theory. II \\
-effects of non-diagonal interactions-    \\
\\
Kazuto Oshima  E-mail address: oshima@nat.gunma-ct.ac.jp　\\
Gunma National College of Technology, Maebashi 371-8530, JAPAN \\
\\
\section*{Abstract}
Spontaneous symmetry breaking in (1+1)-dimensional 
$\phi^{4}$ theory is studied with discretized light-front quantization.   
Taking  effects of non-diagonal interactions into account,
the first few terms  of the commutation relations $[a_{0},a_{n}]$ are recalculated in the $\hbar$ expansion.  Our result of the critical coupling is still consistent with the equal-time result $22\mu^{2}/\hbar \le \lambda_{\rm{cr}} \le 55.5\mu^{2}/\hbar$.  We also have examined effects of regarding the ratio of the bare coupling constant to a renormalized mass as an independent parameter in the $\hbar$ expansion.\\

\section{Introduction}
Light-front QCD appears to be a hope of understanding hadrons from a
field-theoretical point of view.$^{1)}$  Light-front field theory(LFT) is expected to be equivalent to the ordinary field theory not only in the perturbative region,$^{2)}$ but also in the non-perturbative 
region. Spontaneous symmetry breaking(SSB) is a typical phenomenon in the non-perturbative region. One of the most important feature of LFT is that the true vacuum is identical to the Fock vacuum.  Therefore it appears that SSB does not  occur in LFT, but it is now widely believed that a rich low-energy structure of the theory can be explained through zero modes of field operators. 
In  LFT, zero modes of some field operators are not independent variables,$^{3)}$ but obey  certain constraint equations and are given by complex functions of non-zero modes.
In the ordinary equal-time formulation, $(1+1)$-dimensional $\phi^{4}$ model exhibits spontaneous breakdown  of $Z_{2}$ symmetry.$^{4)}$ It is therefore an important issue to confirm that the phase transition of $\phi^{4}$ model is treated correctly in the light-front formulation.
Several authors$^{5)-8)}$ have investigated  the phase transition of the $\phi^{4}$ model in the light-front formulation. In this formulation a non-zero vacuum expectation value of the zero mode is a signal of the phase transition.
In the previous paper$^{9)}$ the author and Yahiro have pointed out
that if we can calculate the commutation relations of the zero mode and the non-zero modes $[a_{0},a_{n}]$,  
we can compute the critical coupling  of the $(1+1)$-dimensional light-front $\phi^{4}$ model. In Ref.~9) we calculated the main part of $[a_{0},a_{n}]$ to $O(\hbar^{4})$ in the $\hbar$ expansion.  We  neglected non-diagonal interactions, which cause  transitions among different modes, since they only give vanishing contributions for certain vacuum expectation values at the first level of the interactions.  In this paper we take  the non-diagonal interactions into account besides diagonal interactions, since the non-diagonal interactions give non-vanishing contributions at the second level of the interactions and subsequently the critical coupling will shift.
Taking all the interactions into account we evaluate the critical coupling within the third-order corrections of the $\hbar$ expansion.  
We also calculate a renormalized mass using front-form pertubation theory and
compute the critical coupling in terms of the renormalized mass. 

This paper is organized as follows: In ${\S}$~2 we denote some conventions of light-front theory and formulate $(1+1)$-dimensional  $\phi ^{4}$ model. In ${\S}$~3 we calculate the first few terms of the commutation relations $[a_{0},a_{n}]$ in the $\hbar$ expansion. In ${\S}$~4 we calculate the vacuum expectation value of the zero mode and compute the critical coupling.
In ${\S}$~5 we calculate the relation between a bare mass and a renormalized mass and evaluate  the  critical coupling in terms of the renormalized mass.  In the last section we give summary.  The text is supplemented by five appendices. In these appendices, several relevant commutation relations are given to calculate $[a_{0},a_{n}]$.  These commutation relations are related each other.

\section{The model and formulation}
We define the light-front coordinates $x^{\pm}=(x^{0}\pm x^{1})/\sqrt{2}$.
The Lagrangian density of  $(1+1)$-dimensional $\phi ^{4}$ theory
is written
\begin{equation}
{\cal L}=\partial_{+}\phi\partial_{-}\phi-{\mu ^{2} \over 2}\phi ^{2}
-{\lambda \over 4!}\phi ^{4}. 
\end{equation}
We put the quantum system in a box of length $d$ and impose periodic boundary
conditions.$^{3),10)}$ The field operator $\phi$ is then expanded as
\begin{equation}
\phi (x)={1 \over \sqrt{4 \pi}}a_{0}+
\sum _{n \ne 0}{1 \over \sqrt {4\pi |n|}}a_{n}(x^{+})e^{ik_{n}^{+}x^{-}},
\end{equation}
where $k_{n}^{+}=2\pi n /d$.  
The coefficients of the expansion are operators which  
satisfy the canonical commutation relations$,^{3),7),8)}$
\begin{equation}
[a_{k},a_{l}]=[a_{k}^{\dagger},a_{l}^{\dagger}]=0, \hspace{5mm}
[a_{k},a_{l}^{\dagger}]=\hbar \delta _{k,l}, \hspace{5mm}
k,l > 0,
\label{commutation}
\end{equation}
where $a_{k}^{\dagger}=a_{-k}$. 
The zero-mode operator $a_{0}$ is not an independent quantity, since 
it governs the constraint equation$^{3),7)}$
\begin{eqnarray}
0&=&a_{0}^{3}+ga_{0}+\sum _{n \geq 1}{1 \over n}(a_{0}a_{n}a_{n}^{\dagger}
+a_{0}a_{n}^{\dagger}a_{n}+a_{n}a_{n}^{\dagger}a_{0}+a_{n}^{\dagger}a_{n}a_{0} \nonumber \\
&&+a_{n}a_{0}a_{n}^{\dagger}+
a_{n}^{\dagger}a_{0}a_{n}- 3\hbar a_{0}) +6{\textstyle \sum _{3}} 
\equiv \Phi (a,a^{\dagger}) ,
\label{constraint}
\end{eqnarray}
where $g \equiv 24\pi \mu^{2}/\lambda$  is the only parameter of the theory and
we have added the term $\sum _{n \geq 1}{-3 \over n}\hbar a_{0}$
to remove tadpole divergences and $ {\textstyle \sum_{n}}$ are defined as
\begin{equation}
{\textstyle \sum_{n}} \equiv {1 \over n!}\sum _{k_{1},k_{2}\cdots ,k_{n} \ne 0}
{\delta_{k_{1}+k_{2}+\cdots +k_{n},0} \over \sqrt{ |k_{1}k_{2}\cdots k_{n}|} }:a_{k_{1}}a_{k_{2}}\cdots a_{k_{n}}:. 
\end{equation}
The commutation relations $(\ref{commutation})$ and 
the constraint equation $(\ref{constraint})$ were obtained  
by quantizing the classical system with the Dirac-Bergmann quantization procedure. This procedure does not specify an operator ordering 
in the constraint equation $(\ref{constraint})$.
Hence we have assumed the Weyl (symmetric) ordering$.^{11)}$  

In the previous paper$^{9)}$ the author and Yahiro have emphasized the importance of the commutation relations $[a_{0},a_{n}]$ in calculating the vacuum expectation value $\langle 0|a_{0}|0 \rangle $. We  obtained the first few terms of $[a_{0},a_{n}]$ in the $\hbar$ expansion, neglecting the cubic term $\sum_{3}$ in $(\ref{constraint})$ which causes the
non-diagonal interactions for simplicity. In fact, at the first level of the interactions, it does not contribute to the vacuum expectation value $ \langle 0|\Phi (a,a^{\dagger})|0 \rangle$. We also used  $[a_{0},a_{m}^{\dagger}a_{m}]=0$, which hold when the cubic term is absent. In this paper we recalculate $  \langle 0|\Phi (a,a^{\dagger}) |0\rangle $ within $O(\hbar^{5})$ without neglecting  the cubic term. The cubic term gives non-zero contributions at the second level of the interactions.

\section{Commutation Relations $[a_{0},a_{n}]$}
In this section we will calculate [$a_{0},a_{n}$] from
\begin{equation} 
[\Phi (a,a^{\dagger}),a_{n}]=0.  
\label{derivation}
\end{equation}
Using the commutation relations ($\ref{commutation}$), Eq.($\ref{derivation}$) is written explicitly as 
\begin{eqnarray}
&&A[a_{0},a_{n}]−{1\over n}a_{0}a_{n}\hbar+{1\over n}[a_{0},a_{n}]\hbar 
−{1\over 6}\sum_{m \geq 1} {1\over m}[[[a_{0},a_{n}],a_{m}],a_{m}^{\dagger}] \nonumber \\
&&−{1\over 2}\sum_{k\ne 0,n}{1\over \sqrt{|nk(n-k)}|}a_{k}a_{n-k}\hbar　+{1\over 2}\sum_{m \geq 1}{1\over m}[[a_{0},a_{m}^{\dagger}a_{m}],a_{n}]\nonumber \\
&&+{1 \over 2}a_{0}[[a_{0},a_{n}],a_{0}]
+{1 \over 6}[[[a_{0},a_{n}],a_{0}],a_{0}]=0,
\label{explicit}
\end{eqnarray}
where
\begin{equation}
A={g \over 6}+{1 \over 2}a_{0}^{2}+\sum _{m \geq 1}{1 \over m}a_{m}^{\dagger}a_{m}. \footnote{The operator $A$ in Ref.~9) is different from $A$ ($\ref{A}$) by a factor 6.}
\label{A}
\end{equation}
In this paper, we will use the letter $m$ as a natural number and the letters $k$ and $l$ as integers. We  mainly use the letter $n$  as a natural number except for some cases, where we use the letter $n$  as a non-zero integer. 
The vacuum expectation value of the fifth term($\sim \sum a_{k}a_{n-k}$) in 
($\ref{explicit}$) which originates from the cubic term $\sum_{3}$ in ($\ref{constraint}$)  vanishes and the last three terms in ($\ref{explicit}$)  vanish or become higher order in $a_{0}$ as long as the  fifth term is neglected($a_{0}$ is very small around the critical point ). In this reason  in the previous paper$^{9)}$ we have neglected the last four terms in ($\ref{explicit}$). 

We will fix $[a_{0},a_{n}]$ order by order in the $\hbar$ expansion. 
At the lowest level we have
\begin{equation}
[a_{0},a_{n}]={1 \over nA}a_{0}a_{n}\hbar+{1 \over 2A}\sum _{k \ne 0,n}{1 \over \sqrt{|kn(n-k)|}}a_{k}a_{n-k}\hbar +O(\hbar^{2}). 
\label{lowest1}
\end{equation}
The right hand side of the commutation relations $[a_{0},a_{n}]$ can be divided into two groups by the powers of the zero mode $a_{0}$.  We call  such terms as the first term in (\ref{lowest1}) diagonal that contain odd powers of $a_{0}$. In the contrast, we call such terms as the second term in (\ref{lowest1}) non-diagonal that contain even powers of $a_{0}$. Note that the powers of $a_{0}$ can increase only through  the commutation relations [${1 \over A},a_{n}$].

Since there is no {\it a priori} way of fixing the operator ordering in ($\ref{lowest1}$), we have fixed the operator ordering so that we can calculate the vacuum expectation value easily. Although the difference of the operator ordering affects the higher-order terms in the expansion, the vacuum expectation values $\langle 0|[a_{0},a_{0}]a_{n}^{\dagger}|0 \rangle$  would not be affected.  
We get [$a_{0},a_{n}^{\dagger}$] by Hermite conjugation of ($\ref{lowest1}$)
\begin{eqnarray}
[a_{0},a_{n}^{\dagger}]&=&-a_{n}^{\dagger}a_{0}{1 \over nA}\hbar-\sum _{k \ne 0,n}{1 \over \sqrt{|kn(n-k)|}}a_{n-k}^{\dagger}a_{k}^{\dagger}{1 \over 2A}\hbar +O(\hbar^{2}) \nonumber \\
&=&-{1 \over nA}a_{0}a_{n}^{\dagger}\hbar-{1 \over 2A}\sum _{k \ne 0,n}{1 \over \sqrt{|kn(n-k)|}}a_{-k}a_{k-n}\hbar +O(\hbar^{2}).　\nonumber \\ 
\label{lowest2}
\end{eqnarray}
Owing to the difference of the operator ordering, the $O(\hbar^{2})$ term in the second line  differs from that in the first line.
We  adopt the operator ordering  in the second line, where $A$ and $a_{0}$ are rearranged in order from the left. By noting $a_{n}^{\dagger}=a_{-n}$ and changing some variables as $k \rightarrow -k$, (\ref{lowest1}) and (\ref{lowest2}) can be combined as
\begin{equation}
[a_{0},a_{n}]={1 \over A}{1 \over n}a_{0}a_{n}\hbar +{1 \over 2A}\sum _{k \ne 0,n}{\epsilon(n) \over \sqrt{|kn(n-k)|}}a_{k}a_{n-k}\hbar+O(\hbar^{2}), \\
 n \ne 0,
\label{lowestcom}
\end{equation}
where $\epsilon(n)=1(n>0),\epsilon(n)=-1(n<0)$. Note that the second term in ($\ref{lowestcom}$) are quadratic in the non-zero modes due to the non-diagonal interactions.  

Before proceeding to higher order in $\hbar$, let us examine what terms should be left in our approximation. We consider the following expansion (see Ref.~9))
\begin{eqnarray}
& &[a_{0},a_{n}] \equiv {\displaystyle \sum _{p \geq 1}}[a_{0},a_{n}]_{p}\hbar^{p} \nonumber \\
&=&{\displaystyle \sum _{p \geq 1}}
{ \alpha _{p}(n) \over A^{p}}a_{0}a_{n} \hbar ^{p}+{\displaystyle \sum _{p \geq 3}}{\displaystyle \sum _{l \geq 1}}
{ \beta_{p,l}(n) \over A^{p+1}}a_{l}^{\dagger}a_{l}a_{0}a_{n} \hbar ^{p} \nonumber \\
&+&{\displaystyle \sum _{p \geq 5}}{\displaystyle \sum _{k,l \geq 1}} 
{ \gamma_{p,kl}(n) \over A^{p+2}}a_{k}^{\dagger}a_{k}a_{l}^{\dagger}a_{l}a_{0}a_{n} \hbar ^{p}
+{1 \over 2A}\sum _{k \ne 0,n}{1 \over \sqrt{|kn(n-k)|}}a_{k}a_{n-k}\hbar \nonumber \\
&+&({\rm higher-order \;  terms})+O(a_{0}^{3}) , \quad n \geq 1, 
\label{expansion}
\end{eqnarray}
where $\alpha _{p}(n),\beta_{p,l}(n)$ and $\gamma_{p,kl}$ are the coefficients
of the diagonal terms to be determined.  
Now we comment on the the expansion ($\ref{expansion}$). 
Since [${1 \over A},a_{l}$]$\ni  {1 \over A^{2}}a_{l}\hbar$,  [$a_{0},a_{l}$]$\ni  {1 \over A}a_{0}a_{l}\hbar$ and the same relations hold for $a_{l}^{\dagger}$,  the terms ${1 \over A^{4}}a_{l}^{\dagger}a_{l}a_{0}a_{n}\hbar^{3}$ in ($\ref{expansion}$) are produced from the fourth term in ($\ref{explicit}$).
  The non-diagonal terms do not directly give non-zero contributions, but they affect the diagonal terms at higher orders, and at O($\hbar^{2}$) we will write down $[a_{0},a_{n}]$ completely.
At  O($\hbar^{3}$) and O($\hbar^{4}$) only the terms ${ \alpha _{p}(n) \over A^{p}}a_{0}a_{n}  \hbar ^{p}(p=3,4)$  give non-zero contributions to the vacuum expectation value $\langle 0|\Phi (a,a^{\dagger}) |0\rangle $  within O($\hbar^{5}$). For example, the terms ${ \beta_{3,l}(n) \over A^{4}}a_{l}^{\dagger}a_{l}a_{0}a_{n} \hbar ^{3}$ can indirectly give non-zero contributions  to   $\langle 0|\Phi (a,a^{\dagger}) |0\rangle $ only at $O(\hbar^{6})$. 
 At O($\hbar^{3}$) and O($\hbar^{4}$) therefore we only pay attention to terms that reduce to $a_{0}a_{n}\hbar^{3}$ or $a_{0}a_{n}\hbar^{4}$. 

At $O(\hbar^{2})$ only four terms in the constraint equation ($\ref{explicit}$) are concerned.  These terms are calculated from $[a_{0},a_{n}]_{1}$ and  $[{1 \over A},a_{n}]_{1}$. The fundamental commutation relations $[{1 \over A},a_{n}]_{1}$ and some secondary commutation relations are denoted in Appendices A, B and C.
From ($\ref{lowestcom}$), (${\rm \ref{mmn1}}$) and  (${\rm \ref{00n0}}$) we obtain 
\begin{eqnarray}
&&[a_{0},a_{n}]_{2} \nonumber \\
&=&-{1 \over n^{2}A^{2}}a_{0}a_{n}+{1 \over 2n^{2}A^{3}}a_{0}^{3}a_{n} \nonumber \\
&&+{\epsilon(n) \over 8A^{3}}\sum_{k \ne 0,n}\sum_{l \ne 0,n-k}{a_{0}(a_{l}a_{n-k-l}a_{k}+a_{k}a_{l}a_{n-k-l}) \over (n-k)\sqrt{|nkl(n-k-l)|}} \nonumber \\
&-&{\epsilon(n) \over 4}\sum_{k \ne 0,n}\Bigl({1 \over n}+{1 \over k}+{1 \over n-k}\Bigr)\Bigl({1 \over A^{2}}-{1 \over A^{3}}a_{0}^{2}\Bigr){1 \over \sqrt{|nk(n-k)|}}a_{k}a_{n-k} \nonumber \\
&-&\sum_{m=1}{\epsilon(m) \over 4mn}\Bigl({1 \over A^{3}}-{1 \over A^{4}}a_{0}^{2}\Bigr)\sum_{k \ne 0,m}{a_{n}(a_{-m}a_{k}a_{m-k}-a_{m}a_{-k}a_{-m+k})   \over \sqrt{|mk(m-k)|}}\nonumber \\
&+&{\epsilon(n) \over 8A^{4}}\sum_{m=1}\sum_{l \ne 0,n}\sum_{k \ne 0,m}{\epsilon(m) \over m\sqrt{|ln(n-l)km(m-k)|}} \nonumber \\
&\times& a_{0}[a_{n-l}a_{l},a_{-m}a_{k}a_{m-k}-a_{m}a_{-k}a_{k-m}],\quad n \ne 0. 
\label{second}
\end{eqnarray}
Although the last term in ($\ref{second}$) belongs to $[a_{0},a_{n}]_{3}$, we write it down here so that we may forget it at $O(\hbar^{3})$.  We have left the symbols $\epsilon(m)(=1)$ as they are, since they are useful to distinguish
the non-diagonal terms from the diagonal terms.    

At $O(\hbar^{3})$ six terms in the constraint equation ($\ref{explicit}$) are concerned.  To obtain $[a_{0},a_{n}]_{3}$ several commutation relations at
$O(\hbar)$ and $O(\hbar^{2})$ are needed. Relevant commutation relations are denoted in Appendices.  Non-diagonal terms 
in $[a_{0},a_{n}]_{3}$ do not give non-zero contributions to the vacuum expectation value $\langle 0|\Phi (a,a^{\dagger}) |0\rangle $ within $O(\hbar^{5})$. 
Neglecting irrelevant terms, from ($\ref{second}$),(${\rm \ref{mmn2}}$),(${\rm \ref{00n0}}$),(${\rm \ref{n000}}$) and (${\rm \ref{0nmm}}$)  we have
\begin{eqnarray}
&&[a_{0},a_{n}]_{3} \nonumber \\
&&=\Bigl({4 \over 3n^{3}}+{\zeta(2) \over 3n}\Bigr){1 \over A^{3}}a_{0}a_{n} \nonumber \\
&&-{1 \over A^{4}}\sum_{k \ne 0,n}\sum_{l \ne 0,n-k}\Bigl({1 \over 6n}+{11 \over 48k}+{5 \over 48l}+{1 \over 8(n-k)}+{5\over 48(n-k-l)}\Bigr) \nonumber \\
&&\times {1 \over (n-k)\sqrt{|nkl(n-k-l)|}}a_{0}(a_{l}a_{n-k-l}a_{k}+a_{k}a_{l}a_{n-k-l}) \nonumber \\
&&+{1 \over 8A^{4}}\sum_{m=1}\sum_{k \ne 0,n,m,m-n}{1 \over m(m-k)\sqrt{|mkn(m-k-n)|}} \nonumber \\
&&\times  a_{0}(a_{-k}a_{n+k-m}+a_{n+k-m}a_{-k})a_{m} \nonumber \\
&&-{1 \over 4A^{4}}\sum_{m=1(m \ne n)}\sum_{k \ne 0,m}{1 \over m^{2}\sqrt{|n(m-n)k(m-k)|}}a_{0}a_{n-m}a_{k}a_{m-k} \nonumber \\
&&-{1 \over 12A^{4}} \sum_{m=1,m\ne n}\sum_{l \ne 0,m}{1 \over m^{2}\sqrt{|ln(m-l)(n-m)|}}a_{0}a_{l}a_{m-l}a_{n-m}\nonumber \\
&&-{1 \over 12A^{4}}\sum_{k \ne 0,n}\sum_{l \ne 0,n-k}{1
\over (n-k)\sqrt{|knl(n-k-l)|}} \nonumber \\
&& \times a_{0}\Bigl({1 \over n-k}a_{l}a_{n-k-l}a_{k}+{1 \over k}a_{k}a_{l}a_{n-k-l}\Bigr) , \quad n>0.
\label{third}
\end{eqnarray}

 We only have to pick up terms including $a_{0}a_{n}\hbar^{4}$ in ($\ref{explicit}$) to calculate $[a_{0},a_{n}]_{4}$. 
From ($\ref{third}$) and (${\rm \ref{0nmm}}$) we have
\begin{equation}
[a_{0},a_{n}]_{4}=-\Bigl({7 \over 3n^{4}}+{5\zeta(2) \over 6n^{2}}+{\zeta(3) \over 2n}\Bigr){1 \over A^{4}}a_{0}a_{n}.
\label{fourth}
\end{equation}
\section{The critical coupling}
Using the commutation relations $[a_{0},a_{n}]$ in the previous section, we calculate the vacuum expectation values $\langle 0|[a_{0},a_{n}]_{4}a_{n}^{\dagger}|0\rangle $ to compute the critical coupling.
Sandwiching ($\ref{lowest1}$) with $\langle 0|$ and $a_{n}^{\dagger}|0 \rangle $, we have 
\begin{equation}
\langle 0|[a_{0},a_{n}]_{1}a_{n}^{\dagger}|0\rangle \hbar 
={6 \over {\tilde g}}{1 \over n}\sigma\hbar^{2},
\end{equation}
where $\sigma \equiv \langle 0|a_{0}|0\rangle$ and ${\tilde g}=g+3\sigma^{2}$.
In the same way from ($\ref{fourth}$) we have
\begin{equation}
\langle 0|[a_{0},a_{n}]_{4}a_{n}^{\dagger}|0\rangle \hbar^{4} 
=-\Bigl({6 \over {\tilde g}}\Bigr)^{4}\Bigl({7 \over 3n^{4}}+{5 \zeta(2) \over 6n^{2}}+{\zeta(3) \over 2n}\Bigr)\sigma\hbar^{5},
\end{equation}
where $\zeta(n)$ is the Riemann zeta function. Using identities such as
$\sum_{k=1}^{n-1}{1 \over k(n-k)}= {2 \over n}\sum_{k=1}^{n-1}{1 \over k}$,
from ($\ref{second}$) and ($\ref{third}$) we have
\begin{eqnarray}
\langle 0|[a_{0},a_{n}]_{2}a_{n}^{\dagger}|0\rangle \hbar^{2} 
&=&-\Bigl({6 \over {\tilde g}}\Bigr)^{2}{1 \over n^{2}}\sigma \hbar^{3}+\Bigl({6 \over {\tilde g}}\Bigr)^{3}\Bigl({1 \over 4n^{3}}+{1 \over 4n^{2}}\sum_{k=1}^{n-1}{1 \over k}\Bigr)\sigma \hbar^{4} \nonumber \\
&+&\Bigl({6 \over {\tilde g}}\Bigr)^{4}\Bigl({1 \over 2n^{2}}\sum_{k=1}^{n-1}{1 \over k^{2}}+{1 \over 2n^{3}}\sum_{k=1}^{n-1}{1 \over k}\Bigr)\sigma \hbar^{5},
\label{v2}
\end{eqnarray}
\begin{eqnarray}
&&\langle 0|[a_{0},a_{n}]_{3}a_{n}^{\dagger}|0\rangle \hbar^{3} 
= \Bigl({6 \over {\tilde g}}\Bigr)^{3}\Bigl({4 \over 3n^{3}}+{\zeta(2) \over 3n}\Bigr)\sigma\hbar^{4} \nonumber \\
&&-\Bigl( {6 \over {\tilde g}}\Bigr)^{4}\Bigl({1 \over 3n^{4}}-{1 \over 6n^{2}}\zeta(2)+{17 \over 6n^{3}}\sum_{k=1}^{n-1}{1 \over k}+{13 \over 12n^{2}}\sum_{k=1}^{n-1}{1 \over k^{2}}\Bigr)\sigma\hbar^{5}. \nonumber \\
\label{v3}
\end{eqnarray}
The $O(\hbar^{4})$ and $O(\hbar^{5})$ terms in ($\ref{v2}$) and $O(\hbar^{5})$ terms in ($\ref{v3}$)  originate from the non-diagonal interactions. 
Collecting the above results, we have
\begin{eqnarray}
&&\langle 0|[a_{0},a_{n}]a_{n}^{\dagger}|0\rangle  
={6 \over {\tilde g}}{1 \over n}\sigma\hbar^{2} -\Bigl({6 \over {\tilde g}}\Bigr)^{2}{1 \over n^{2}}\sigma\hbar^{3} \nonumber \\
&+&\Bigl({6 \over {\tilde g}}\Bigr)^{3}\Bigl({4 \over 3n^{3}}+{\zeta(2) \over 3n} 
+{1 \over 4n^{3}}+{1 \over 4n^{2}}\sum_{k=1}^{n-1}{1 \over k}\Bigr)\sigma\hbar^{4} \nonumber \\
&-&\Bigl({6 \over {\tilde g}}\Bigr)^{4}\Bigl({7 \over 3n^{4}}+{5\zeta(2) \over 6n^{2}}+{\zeta(3) \over 2n}+{1 \over 3n^{4}}-{\zeta(2) \over 6n^{2}} \nonumber \\
&+&{7 \over 3n^{3}}\sum_{k=1}^{n-1}{1 \over k}+{7 \over 12n^{2}}\sum_{k=1}^{n-1}{1 \over k^{2}}\Bigr)\sigma\hbar^{5}.
\label{0n}
\end{eqnarray}
Sandwiching ($\ref{constraint}$) with the Fock vacuum $|0\rangle$, we have
\begin{eqnarray}
&0&=\sigma^{3}+g\sigma -\sum_{n=1}^{\infty}{1 \over n}\langle 0|[a_{0},a_{n}]a_{n}^{\dagger}|0\rangle  \nonumber \\
&&=\sigma^{3}+g\sigma\nonumber \\
&&-\biggl[{6 \over {\tilde g}}\zeta(2)\hbar^{2}-\Bigl({6 \over {\tilde g}}\Bigr)^{2}\zeta(3)\hbar^{3}+\Bigl({6 \over {\tilde g}}\Bigr)^{3}\Bigl({19 \over 12}\zeta(4) +{1 \over 3}(\zeta(2))^{2}+\sum_{n=1}^{\infty}{1 \over 4n^{3}}\sum_{k=1}^{n-1}{1 \over k}\Bigr)\hbar^{4}  \nonumber \\
&&-\Bigl({6 \over {\tilde g}}\Bigr)^{4}\Bigl({8 \over 3}\zeta(5)+{7 \over 6}\zeta(3)\zeta(2)+\sum_{n=1}^{\infty}{7 \over 3n^{4}}\sum_{k=1}^{n-1}{1 \over k}+\sum_{n=1}^{\infty}{7 \over 12n^{3}}\sum_{k=1}^{n-1}{1 \over k^{2}}\Bigr)\hbar^{5}\biggr]\sigma.\nonumber
\\
\label{constraint2}
\end{eqnarray}
At the critical point non-trivial solutions start to appear besides the trivial
solution $\sigma =0$.
From ($\ref{constraint2}$) the critical point is given by the condition that   the linear terms in $\sigma$ balance to zero.　This condition is written  as
\begin{equation}
0=36-x^{2}(1-0.2326x+0.1665x^{2}-0.1068x^{3}),
\label{bcritical}
\end{equation}
where $x \equiv 6\pi\hbar/g=\lambda \hbar/4\mu^{2}$ and we have substituted the values $\sum_{n=1}^{\infty}{1 \over n^{3}}\sum_{k=1}^{n-1}{1 \over k}=0.3529$,$\sum_{n=1}^{\infty}{1 \over n^{3}}\sum_{k=1}^{n-1}{1 \over k^{2}}=0.2288,$\\
$\sum_{n=1}^{\infty}{1 \over n^{4}}\sum_{k=1}^{n-1}{1 \over k^{2}}=0.0966$. Owing to the non-diagonal interactions, the third and the fourth terms in the parentheses in ($\ref{bcritical}$) have increased by some 15$\%$ and 8$\%$ respectively over the previous results in Ref.~9), where the non-diagonal interactions are neglected.  

We adopt the Pad${\rm {\acute e}}$ approximations$^{12)}$ to the alternating 
series $1-c_{1}x+c_{2}x^{2}-c_{3}x^{3}+\cdots$. This series is approximated
as $[M/N]$, which denotes a rational equation of polynomials of degrees $M$ and $N$:
\begin{eqnarray}
&[i/1]& \qquad \qquad 1-c_{1}x+\cdots +(-1)^{i}c_{i}{x^{i} \over 1+{c_{i+1} \over c_{i}}x} ,   \\
&[1/2]&  \qquad \qquad  {1+({c_{3}-c_{1}c_{2} \over c_{2}-c_{1}^{2}}-c_{1})x
\over 1+{c_{3}-c_{1}c_{2} \over c_{2}-c_{1}^{2}}x+({c_{3}-c_{1}c_{2} \over c_{2}-c_{1}^{2}}c_{1}-c_{2})x^{2}}.
\end{eqnarray}
Our results are 
\begin{eqnarray}
&[0/0]& \quad \lambda_{{\rm cr}}=24 {\mu}^{2}/\hbar , \qquad \; \>
[1/1] \quad  \lambda_{{\rm cr}}=28.1{\mu}^{2}/\hbar, \nonumber \\ 
&[0/1]& \quad  \lambda_{{\rm cr}}=46.0 {\mu}^{2}/\hbar, \qquad 
[2/1] \quad  \lambda_{{\rm cr}}=26.0 {\mu}^{2}/\hbar ,\qquad \nonumber \\ &[1/2]& \quad  \lambda_{{\rm cr}}=25.5 {\mu}^{2}/\hbar.
\label{valx}
\end{eqnarray}
In Ref.~9) we have assumed that the alternating series has such properties that the lower(upper) bounds of  $\lambda _{{\rm cr}}$ are obtained from cases $M+N$=even(odd). The values in ($\ref{valx}$) exhibit that this assumption does not hold.   However, all the values in  ($\ref{valx}$) still lie within the range $22 {\mu}^{2}/\hbar \le \lambda _{{\rm cr}} \le 55.5 {\mu}^{2}/\hbar$ of the equal-time result.

The ratio of the bare coupling constant to the bare mass is the only parameter
of the theory and in the literature $g_{{\rm cr}}$ is computed. In two dimension the coupling constant and the mass suffer only finite quantum corrections  in this model.  However, in four dimension the bare quantities become infinite and it will be adequate to describe the theory by renormalized quantities. Hence as a preliminary to study four-dimensional models, we adopt the ratio of the bare coupling constant to a renormalized mass as a new parameter in the next section.
We avoid treating coupling constant renormalization since it seems not so easy 
to  calculate quantum corrections to sufficient order in $\hbar$.
In the next section we compute  the critical value of the bare coupling constant in terms of the renormalized mass.   
\section{The critical coupling in terms of the renormalized mass}
 Let us  study a relation between the bare mass $\mu$ and the renormalized mass $m$ for the model.
We split the Hamiltonian into two parts $H={d\mu^{2} \over 4\pi}(H_{0}+H_{I})$, where $H_{0}=g\sum_{2}=g\sum_{n=1}^{\infty}{a_{n}^{\dagger}a_{n} \over n}$ and $H_{I}=6\sum_{4}+H_{ZM}$. The Hamiltonian $H_{ZM}$ contains all of the zero-mode interactions 
\begin{eqnarray}
H_{ZM}&=&-{a_{0}^{4} \over 4}+{1 \over 4}\sum_{n \ne 0}{1 \over |n|}(a_{n}a_{0}^{2}a_{-n}-a_{0}a_{n}a_{-n}a_{0})\nonumber
\\  &+&{1 \over 4}\sum_{k,l,n \ne 0}{\delta_{k+l+n,0} \over \sqrt{|kln|}}(a_{k}a_{0}a_{l}a_{n}+a_{k}a_{l}a_{0}a_{n}).
\label{zeromode}
\end{eqnarray}
The states $a_{p}^{\dagger}|0 \rangle $ are eigenstates of the unperturbed Hamiltonian $H_{0}$ with eigenvalues ${g \over p}\hbar={24\pi \mu^{2} \over p\lambda}\hbar$.　We approximately define the renormalized mass
$m$ through  eigenvalues of $H$ as 
\begin{equation}
H a_{p}^{\dagger}|0 \rangle \approx {24\pi m^{2} \over p\lambda}\hbar a_{p}^{\dagger}|0 \rangle .
\label{rmass}
\end{equation}
If the interactions were absent we would obtain the bare mass $\mu$, hence this  definition of the renormalized mass $m$ will be plausible. We will solve the eigenvalue equation ($\ref{rmass}$) perturbatively. 

Effects of the interaction $H_{I}$ have already been calculated to the second order in  perturbation theory in the limit $d \rightarrow \infty$ in Ref.~8).   
Taking ($\ref{bcritical}$) into consideration, we calculate effects of the interaction $H_{I}$  to the third order in  perturbation theory in the limit $d \rightarrow \infty$ .

First, we calculate the contributions of the non-zero modes.  Since we have added the counter term in  ($\ref{constraint}$), the tad pole diagrams always vanish.  At the second order in perturbation theory we have a non-vanishing diagram (see Fig.$~1$):  
\begin{eqnarray}
& &\langle 0 |a_{p}(6{\textstyle \sum{}_{4}}){1 \over E-g\sum_{2}}(6{\textstyle \sum{}_{4}})a^{\dagger}_{p}|0\rangle \nonumber \\ 
&=&{6 \over gp}\sum_{i,j>0}^{i+j<p}{1 \over ip(p-i-j)\big({1 \over p}-{1 \over i}-{1 \over j}-{1 \over p-i-j}\big)}\hbar^{4} \nonumber \\ 
&\longrightarrow& -{6 \over gp}\int_{0}^{P}dx\int_{0}^{P-x}dy{P \over (x+y)(P-x)(P-y)}\hbar^{4} 
=-{6 \over gp}{\pi^{2} \over 4}\hbar^{4}, \nonumber \\
\label{m2}
\end{eqnarray}
where $i,j$ and $p$ are natural numbers and $E={g \over p}\hbar$ is the unperturbed energy;
we have set  $x=2\pi i/d,y=2\pi j/d$ and $P=2\pi p/d$ and have taken the limit $d,p \rightarrow \infty$ so that $P$ remains finite. Equation ($\ref{bcritical}$) is independent of $d$ and it holds also at large $d$. Therefore in the following
we consider the situation $d \rightarrow \infty$.  
In the ordinary equal-time formulation the
corresponding Feynman diagram of  Fig.$~1$ gives the same result.$^{8)}$
In the same way at the third order in perturbation theory(see  Fig.$~2$) we have
\begin{equation}
\langle 0 |a_{p}(6{\textstyle \sum{}_{4}}){1 \over E-g\sum_{2}}(6{\textstyle \sum_{4}}){1 \over  E-g\sum_{2}}(6{\textstyle \sum_{4}})a^{\dagger}_{p}|0\rangle 
={54 \over pg^{2}}I\hbar^{5},
\label{m3}
\end{equation}
where  $I$ is the following quantity which approaches a definite value in the limit $d \rightarrow \infty$
\begin{eqnarray}
&&I= \nonumber \\
&&\sum_{i,j,k>0}^{k<i+j<p}{1 \over ip(p-i-j)\big({g \over p}-{1 \over i}-{1 \over j}-{1 \over p-i-j}\big) k(i+j-k)\big({g \over p}-{1 \over k}-{1 \over i+j-k}-{1 \over p-i-j}\big)}\nonumber \\ 
&&\longrightarrow \nonumber \\
&&\int_{0}^{P}dx \int_{0}^{P-x}dy\int_{0}^{x+y}dz{P^{2}(P-x-y) \over (x+y)^{2}(P-x)(P-y)(P-x-y+z)}
= 1.645. \nonumber \\
\label{m4}
\end{eqnarray}
\begin{picture}(300,170)
\put(10,100){\line(1,0){100}}
\put(15,92){p}
\put(60,122){i}
\put(60,92){j}
\put(53,72){p-i-j}
\put(100,92){p}
\put(60,100){\circle{40}}
\put(190,100){\line(1,0){30}}
\put(240,100){\circle{40}}
\put(260,100){\line(1,0){30}}
\put(220,100){\line(1,1){20}}
\put(240,120){\line(1,-1){20}}
\put(195,92){p}
\put(230,120){i}
\put(230,102){j}
\put(246,102){k}
\put(251,120){i+j-k}
\put(233,72){p-i-j}
\put(280,92){p}
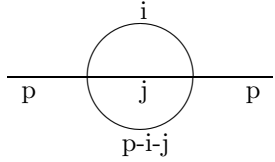
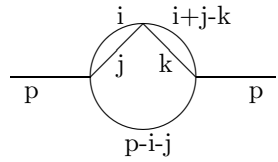
\put(0,40){Fig.~1.  The second order diagram.}  
\put(180,40){Fig.~2.  The third order diagram.}
\end{picture}
\\ 
Second, we calculate the contributions of $H_{zm}$.
Solving  the constraint equation ($\ref{constraint}$) perturbatively in $1/g$ for $a_{0}$, we have
\begin{equation}
a_{0}=-{6 \over g}{\textstyle \sum_{3}}+{6 \over g^{2}}\sum_{n \ne 0}
\bigl({\textstyle \sum_{3}}a_{n}a_{-n}+a_{n}a_{-n}{\textstyle \sum_{3}}+
 a_{n}{\textstyle \sum_{3}}a_{-n} -{3 \over 2}{\textstyle \sum_{3}}\bigr)+\cdots .
\end{equation}
Substituting this expression into $H_{zm}$ ($\ref{zeromode}$), to $O(\hbar^{5})$ we find
\begin{eqnarray}
&&\langle 0 |a_{p}H_{zm}a^{\dagger}_{p}|0\rangle 
=-{3 \over gp}\biggl(\sum_{k=1}^{p-1}{1 \over k(p-k)}+4\sum_{k>0}{1 \over k(p+k)}\biggr)\hbar^{4} \nonumber \\
&&+{18 \over g^{2}p}\biggl(\sum_{k=1}^{p-1}{1 \over k^{2}(p-k)}+{4 \over p}\sum_{k=1}^{p-1}{1 \over k(p-k)}\nonumber \\
&&+\sum_{k>0}{1 \over k^{2}(p+k)}+\sum_{k>0}{1 \over k(p+k)^{2}}-{1 \over p}\sum_{k>0}{1 \over k(p+k)}\biggr)\hbar^{5} ,
\end{eqnarray}
where we have neglected some constant terms, which means a shift of the vacuum energy. Note that $\sum_{k=1}^{p-1}{1 \over k(p-k)}$ and $\sum_{k>0}{1 \over k(p+k)}$ in the $O(\hbar^{4})$ terms vanish in the limit $p \rightarrow \infty$.  The $O(\hbar^{5})$ terms also vanish in this limit. Thus the mass will not be corrected  by  the zero mode in this limit. However, the convergences in ($\ref{m2}$) and ($\ref{m4}$) will be improved by inclusions of the zero mode.$^{8)}$ 

The eigenvalues of $H_{0}$ and their corrections ($\ref{m2}$) and ($\ref{m3}$)  have the common  factor $1/p$ as expected, and we get the following relation between the renormalized mass $m$ and the bare mass $\mu$  
\begin{eqnarray}
m^{2}&=&\mu^{2}\bigl(1-{6 \over g^{2}}{\pi^{2} \over 4}\hbar^{2}
+{54 \over g^{3}}I\hbar^{3}\bigr) +O(\hbar^{4}) \nonumber \\
&=& \mu^{2}\bigl(1-{14.80 \over g^{2}}\hbar^{2}
+{88.83 \over g^{3}}\hbar^{3}\bigr) +O(\hbar^{4}).
\label{m-mu}
\end{eqnarray}
Now we redefine the parameter $g$ as $g=24\pi m^{2}/\lambda$, which we regard
as independent of $\hbar$, instead of  $g=24\pi \mu^{2}/\lambda$.  In the following $g$ means this new parameter.   From ($\ref{m-mu}$) we have
\begin{equation}
{24\pi \mu^{2} \over \lambda}=g+{14.80 \over g}\hbar^{2}-{88.83 \over g^{2}}\hbar^{3}+O(\hbar^{4}).
\end{equation} 

Taking this $\hbar$ dependence into account, the constraint equation ($\ref{explicit}$)  becomes
\begin{equation}
{\rm l.h.s. \; of \;}(\ref{explicit})+{14.80 \over 6g}[a_{0},a_{n}]\hbar^{2}-{88.83 \over 6g^{2}}[a_{0},a_{n}]\hbar^{3}=0.
\end{equation}
These two terms affect  $[a_{0},a_{n}]_{2}$ and $[a_{0},a_{n}]_{3}$
\begin{equation}
[a_{0},a_{n}]_{3}={\rm r.h.s. \; of \;}(\ref{explicit})-{14.80 \over 6gA^{2}}{1 \over n}a_{0}a_{n},
\end{equation}
\begin{equation}
[a_{0},a_{n}]_{4}={\rm r.h.s. \; of \;}(\ref{explicit})+{14.80 \over 3gA^{3}}{1 \over n^{2}}a_{0}a_{n}+{88.83 \over 6g^{2}A^{2}}{1 \over n}a_{0}a_{n}.
\end{equation}
Thus the fluctuations in the $\hbar$ expansion of $[a_{0},a_{n}]$ have diminished in terms of the renormalized mass. 
This time we have instead of ($\ref{bcritical}$)
\begin{equation}
0=36-x^{2}(1-0.23261x+0.124889x^{2}-0.0742x^{3}),
\label{bcriticalm}
\end{equation}
with $x={\lambda \hbar \over 4m^{2}}$. 

Using again the Pad${\rm{\acute e}}$ approximations, this time we have
\begin{eqnarray}
&[0/0]& \quad \lambda_{{\rm cr}}=24 m^{2}/\hbar , \qquad \; \>
[1/1] \quad  \lambda_{{\rm cr}}=29.7m^{2}/\hbar , \nonumber \\ 
&[0/1]& \quad  \lambda_{{\rm cr}}=46.0 m^{2}/\hbar, \qquad 
[2/1] \quad  \lambda_{{\rm cr}}=33.3 m^{2}/\hbar  ,\qquad \nonumber \\ &[1/2]& \quad  \lambda_{{\rm cr}}=32.2 m^{2}/\hbar.
\label{val}
\end{eqnarray}
The variance of the values in ($\ref{val}$) is smaller than that of the values in ($\ref{valx}$).  The values in ($\ref{val}$) seem to approach their mean value as the order $M+N$ of the Pad${\rm{\acute e}}$ approximations increases unlike the case ($\ref{valx}$).

\section{Summary}
Taking  the non-diagonal interactions into account besides the diagonal interactions, we have calculated  the relevant terms of $[a_{0},a_{n}]$ to $O(\hbar^{4})$. In addition to the Riemann zeta function $\zeta(s)=\sum_{n=1}^{\infty}{1 \over n^{s}}$, the more complicated function  $\sum_{n=1}^{\infty}{1 \over n^{s}}\sum_{k=1}^{n-1}{1 \over k^{t}}$ has appeared corresponding to the non-diagonal interactions.
Owing to the non-diagonal interactions, the third and fourth coefficients of the alternating series in ($\ref{bcritical}$) have increased by some 15${\%}$ and 8${\%}$.  We have applied the Pad${\rm{\acute e}}$ approximations to the alternating series to compute the critical coupling.  Although the critical coupling  shifts sensitively to the coefficients in the $\hbar$ expansion, at each order of the Pad${\rm{\acute e}}$ approximations $\lambda_{{\rm cr}}$ lies within the range  $22\mu^{2}/\hbar \le \lambda_{{\rm cr}} \le  55.5\mu^{2}/\hbar$ of the equal-time result.
We also have calculated the renormalized mass in front-form perturbation theory. We have obtained another $\hbar$ expansion of $[a_{0},a_{n}]$ by regarding the ratio of the bare coupling constant to the renormalized mass as a new independent parameter. We have found that the fluctuations in the $\hbar$ expansion of $[a_{0},a_{n}]$  and the variance of the values of the critical coupling have diminished.  
   
\section*{Acknowledgments}
The author would like to thank Dr.M.Yahiro, who invited him to study this subject. 

\section*{Appendix A　$\qquad$ {\rm [${1 \over A},a_{n}$]}}
In this appendix we calculate the fundamental commutation relations [${1 \over A},a_{n}$].  We can obtain [${1 \over A},a_{m}$] from [$a_{0},a_{m}$] as follows
\begin{eqnarray}
\Bigl[{1 \over A},a_{m}\Bigr]
&=&{1 \over A}[a_{m},A]{1 \over A} \nonumber \\
&=&{1 \over A}\Bigl([a_{m},{1 \over 2}a_{0}^{2}]+{\hbar \over m}a_{m}\Bigr){1 \over A} \nonumber \\
&=&\Bigl({\hbar \over m}{1 \over A^{2}}-\Bigl({\hbar \over m}\Bigr)^{2}{1 \over A^{3}}+\Bigl({\hbar \over m}\Bigr)^{3}{1 \over A^{4}}+\cdots \Bigr)a_{m}-{1 \over mA^{3}}a_{0}^{2}a_{m}\hbar \nonumber \\
&-&{1\over 2A^{3}}\sum _{k \ne 0,m}{1 \over \sqrt{|km(m-k)|}}a_{k}a_{m-k}a_{0}\hbar+O(\hbar^{2}), m>0 . \nonumber \\
\label{A+}
\end{eqnarray}
The abbreviated terms lead to higher-order contributions.
In the same way we have
\begin{eqnarray}
\Bigl[{1 \over A},a_{m}^{\dagger}\Bigr]&=&-\Bigl({\hbar \over m}{1 \over A^{2}}+\Bigl({\hbar \over m}\Bigr)^{2}{1 \over A^{3}}+\Bigl({\hbar \over m}\Bigr)^{3}{1 \over A^{4}}+\cdots \Bigr)a_{m}^{\dagger}+ {1 \over mA^{3}}a_{0}^{2}a_{m}^{\dagger}\hbar \nonumber \\
&+&{1\over 2A^{3}}\sum _{k \ne 0,m}{1 \over \sqrt{|km(m-k)|}}a_{m-k}^{\dagger}a_{k}^{\dagger}a_{0}\hbar+O(\hbar^{2}), m>0. \nonumber \\
\label{A-}
\end{eqnarray}
Equations  (${\rm \ref {A+}}$) and (${\rm \ref {A-}}$) are combined as
\begin{eqnarray}
\Bigl[{1 \over A},a_{n}\Bigr]&=&\Bigl({\hbar \over n}{1 \over A^{2}}-\Bigl({\hbar \over n}\Bigr)^{2}{1 \over A^{3}}+\Bigl({\hbar \over n}\Bigr)^{3}{1 \over A^{4}}+\cdots \Bigr)a_{n}-{1 \over nA^{3}}a_{0}^{2}a_{n}\hbar  \nonumber \\
&-&{\epsilon(n) \over 2A^{3}}\sum _{k \ne 0,n}{1 \over \sqrt{|kn(n-k)|}}a_{k}a_{n-k}a_{0}\hbar+O(\hbar^{2}), n \ne 0. \nonumber \\
\label{Acombined}
\end{eqnarray}
\section*{Appendix B $\qquad${\rm [$a_{0},a_{m}^{\dagger}a_{m}$]}}  
In this appendix we calculate the first few terms of $[a_{0},a_{m}^{\dagger}a_{m}]$ in the $\hbar$ expansion.
In  our approximation it is sufficient to get $[a_{0},a_{m}^{\dagger}a_{m}]_{p}(p=1,2,3)$.　
More precisely, we need  the coefficients of the terms in the  expansion  
\begin{equation}
[a_{0},a_{m}^{\dagger}a_{m}]=(a_{0}aa\hbar+a_{0}aa\hbar^{2}+a_{0}aaaa\hbar^{2}+a_{0}aa\hbar^{3})+(aaa\hbar),
\label{mmexp}
\end{equation}
where the parentheses denote the disregard of the coefficients and $a$'s  represent the non-zero modes. 
We have neglected terms such as $(a_{0}^{3}aa\hbar^{2})$ and $(aaa\hbar^{2})$ that become higher order in $\hbar$.

Let us examine how  $[a_{0},a_{m}^{\dagger}a_{m}]$ would be if  the non-diagonal interactions were absent.   At the lowest level we have
\begin{eqnarray}
 [a_{0},a_{m}^{\dagger}a_{m}]_{1} 
&=& a_{m}^{\dagger}[a_{0},a_{m}]_{1}+[a_{0},a_{m}^{\dagger}]_{1}a_{m} \nonumber \\
&\sim& a_{m}^{\dagger}{1 \over A}a_{0}a_{m}\hbar-a_{m}^{\dagger}a_{0}{1 \over A}a_{m}\hbar = a_{m}^{\dagger}\Bigl[{1 \over A},a_{0}\Bigr]a_{m}\hbar \nonumber \\
&\sim& \sum_{l} a_{m}^{\dagger}[a_{l}^{\dagger}a_{l},a_{0}]a_{m}\hbar.
\label{nondiag}
\end{eqnarray}
This equation will indicate that if we neglect the non-diagonal
interactions, $[a_{0},a_{m}^{\dagger}a_{m}]$ can  vanish  at any order of $\hbar$. In the following we calculate the coefficients of (${\rm \ref{mmexp}}$) taking the non-diagonal interactions into account.

First, we calculate the contributions of the diagonal parts  $[a_{0},a_{m}]_{{\rm d}p}$ \\
$(p=1,2,3)$. Using ($\ref{lowestcom}$) and ($\ref{second}$), we have
\begin{eqnarray}
[a_{0},a_{m}^{\dagger}]_{{\rm d}1}a_{m}\hbar &+& a_{m}^{\dagger}[a_{0},a_{m}]_{{\rm d}1}\hbar 
={2 \over A^{2}m^{2}}a_{0}a_{m}^{\dagger}a_{m}\hbar^{2}+{3 \over A^{3}m^{3}}a_{0}a_{m}^{\dagger}a_{m}\hbar^{3}　\nonumber \\
&\oplus &(a_{0}^{3}aa\hbar^{2}+aaa\hbar^{2}+a_{0}^{2}aaa\hbar^{2}+a_{0}aaaa\hbar^{3}). 
\label{dd1}
\end{eqnarray}
Here the symbol $\oplus$  means that the following terms can be neglected in our approximation.  In the following we usually neglect such higher-order terms
that become $O(\hbar^{6})$ in $\langle 0|[a_{0},a_{n}]a_{n}^{\dagger}|0\rangle$.
The $O(\hbar)$ terms have canceled, since $-{1 \over m}+{1 \over m}=0$.   In the same way we have   
\begin{eqnarray}
& &[a_{0},a_{m}^{\dagger}]_{{\rm d}2}a_{m}\hbar^{2}+
a_{m}^{\dagger}[a_{0},a_{m}]_{{\rm d}2}\hbar^{2} \nonumber \\
&=&-{2 \over A^{2}m^{2}}a_{0}a_{m}^{\dagger}a_{m}\hbar^{2}-{3 \over 
A^{3}m^{3}}a_{0}a_{m}^{\dagger}a_{m}\hbar^{3}
 \nonumber \\
&+&{1 \over 8A^{3}}\sum_{k \ne 0,m}\sum_{l \ne 0,n-k}{1 \over (m-k)\sqrt{|mkl(m-l-k)|}} a_{0}(a_{m}^{\dagger}a_{l}a_{m-k-l}a_{k} \nonumber \\
&+&a_{m}^{\dagger}a_{k}a_{l}a_{m-k-l}+a_{l}^{\dagger}a_{m-k-l}^{\dagger}a_{k}^{\dagger}a_{m}+a_{k}^{\dagger}a_{l}^{\dagger}a_{m-k-l}^{\dagger}a_{m})\hbar^{2}. 
\end{eqnarray}
The last term derives from the first $\epsilon^{2}$-term in ($\ref{second}$).
At $O(\hbar^{3})$ we only have to pay attention to the terms including
$a_{0}a_{m}^{\dagger}a_{m}\hbar^{3}$.  Since $-{1 \over m}+{1 \over m}=-{1 \over m^{3}}+{1 \over m^{3}}=0$, from the first term of ($\ref{third}$) we have
\begin{eqnarray}
[a_{0},a_{m}^{\dagger}]_{{\rm d}3}a_{m}\hbar^{3}+
a_{m}^{\dagger}[a_{0},a_{m}]_{{\rm d}3}\hbar^{3}=0\hbar^{3} \oplus (a_{0}aaaa\hbar^{3}).
\end{eqnarray}
   
Second, we calculate the contributions of the non-diagonal parts  $[a_{0},a_{m}^{\dagger}a_{m}]_{{\rm nd}p}(p=1,2)$.
At the lowest level we have
\begin{eqnarray}
&&[a_{0},a_{m}^{\dagger}]_{{\rm nd}1}a_{m}\hbar+a_{m}^{\dagger}[a_{0},a_{m}]_{{\rm nd}1}\hbar \nonumber \\
&=&a_{m}^{\dagger}{1 \over 2A}\sum_{k \ne 0,m}{1 \over \sqrt{|km(m-k)|}}a_{k}a_{m-k}\hbar \nonumber \\
&-&{1 \over 2A}\sum_{k \ne 0,m}{1 \over \sqrt{|km(m-k)|}}a_{-k}a_{-m+k}a_{m}\hbar \nonumber \\
&=&{\epsilon(m) \over 2A}\sum_{k \ne 0,m}{1 \over \sqrt{|km(m-k)|}}(a_{m}^{\dagger}a_{k}a_{m-k}-a_{-k}a_{k-m}a_{m})\hbar \nonumber \\
&-&{\epsilon^{2}(m) \over 4A^{3}}\sum_{l \ne 0,m}\sum_{k \ne 0,m}{1 \over m\sqrt{|l(m-l)k(m-k)|}}a_{0}a_{-l}a_{l-m}a_{k}a_{m-k}\hbar^{2} \nonumber \\
&\oplus &
{1 \over 2}\Bigl({1 \over mA^{2}}+{\hbar \over m^{2}A^{3}}+\cdots \Bigr)
\sum_{k \ne 0,m}{\epsilon(m) \over \sqrt{|km(m-k)|}}a_{m}^{\dagger}a_{k}a_{m-k}\hbar^{2}. \nonumber \\
\label{nd1}
\end{eqnarray}
Since $[{1 \over A}aaa\hbar^{p},a_{n}] = (aa\hbar^{p+1})+ (a_{0}aaaaa\hbar^{p+1})$, the contributions of the first and third terms in (${\rm \ref{nd1}}$) to $\langle 0|[a_{0},a_{n}]a_{n}^{\dagger}|0\rangle $ are $O(\hbar^{5})$ and $O(\hbar^{6})$ respectively.  The second term in (${\rm \ref{nd1}}$) gives non-zero contributions to $[a_{0},a_{m}a_{m}^{\dagger}]_{{\rm d}3}$. Hence we cannot neglect the second term.  
In the same way we have
\begin{equation}
[a_{0},a_{m}^{\dagger}]_{{\rm nd}2}a_{m}\hbar^{2}+a_{m}^{\dagger}[a_{0},a_{m}]_{{\rm nd}2}\hbar^{2}=  \oplus (aaa\hbar^{2}). \nonumber \\
\label{nd2}
\end{equation}
The right hand side of (${\rm \ref{nd2}}$) gives only  higher-order contributions.
Thus as concerned with the non-diagonal interactions, we only have to take the lowest-order terms into account .

Collecting the above results for the diagonal and the non-diagonal interactions, we obtain the main result in this appendix:
\begin{eqnarray}
&&[a_{0},a_{m}^{\dagger}a_{m}] \nonumber \\
&=&{1 \over 2A}\sum_{k \ne 0,m}{1 \over \sqrt{|km(m-k)|}}(a_{m}^{\dagger}a_{k}a_{m-k}-a_{m}a_{k}^{\dagger}a_{m-k}^{\dagger})\hbar \nonumber \\
&+&{1 \over 8A^{3}}\sum_{k \ne 0,m}\sum_{l \ne 0,n-k}{1 \over (m-k)\sqrt{|mkl(m-l-k)|}} \nonumber \\
&\times& a_{0}(a_{m}^{\dagger}a_{l}a_{m-k-l}a_{k}+a_{m}^{\dagger}a_{k}a_{l}a_{m-k-l} \nonumber \\
&&+a_{l}^{\dagger}a_{m-k-l}^{\dagger}a_{k}^{\dagger}a_{m}+a_{k}^{\dagger}a_{l}^{\dagger}a_{m-k-l}^{\dagger}a_{m})\hbar^{2} \nonumber \\ 
 &-&{1 \over 4A^{3}}\sum_{k \ne 0,m}\sum_{l \ne 0,m}{1 \over m\sqrt{|lk(m-l)(m-k)|}}a_{0}a_{-l}a_{-m+l}a_{k}a_{m-k}\hbar^{2}.　\nonumber \\
\label{mm}
\end{eqnarray}
Using (${\rm \ref{mm}}$), we can write down the lowest term
of the expansion of $[[a_{0},a_{m}a_{m}^{\dagger}],a_{n}]$ as\\
\begin{eqnarray}
&&[[a_{0},a_{m}a_{m}^{\dagger}]_{1},a_{n}]_{1}\hbar^{2} \nonumber \\
&=&-{\epsilon(n) \over 4A^{3}}\sum_{l \ne 0,n}\sum_{k \ne 0,m}{\epsilon(m) \over \sqrt{|ln(n-l)km(m-k)|}} \nonumber \\
&\times&a_{0}a_{n-l}a_{l}(a_{m}^{\dagger}a_{k}a_{m-k}-a_{m}a_{k}^{\dagger}a_{m-k}^{\dagger})\hbar^{2} \nonumber \\
&+&{\epsilon(n) \over 2}\sum_{k \ne 0,n}\Bigl(-{1 \over n}+{1 \over k}+{1 \over n-k}\Bigr){1 \over A^{2}}{1 \over \sqrt{|nk(n-k)|}}a_{k}a_{n-k}\hbar^{2} \nonumber \\
&+&{1 \over 2n}\Bigl({1 \over A^{3}}-{1 \over A^{4}}a_{0}^{2}\Bigr)\sum_{k \ne 0,m}{\epsilon(m) \over \sqrt{|mk(m-k)|}}a_{n}(a_{m}^{\dagger}a_{k}a_{m-k}-a_{m}a_{k}^{\dagger}a_{m-k}^{\dagger})\hbar^{2}, \nonumber \\
&& \qquad \qquad \qquad \qquad \qquad \qquad \qquad \qquad \qquad  \qquad \qquad  n \ne 0.
\label{mmn1}
\end{eqnarray}
The first term becomes $O(\hbar^{5})$  when sandwiched between $\langle 0|$ and $a_{n}^{\dagger}|0\rangle$. The second term leads to $O(\hbar^{5})$ contributions through $a_{0}[[a_{0},a_{n}],a_{0}]$(see (${\rm \ref{a021}}$)). The next higher-order term $[[a_{0},a_{m}a_{m}^{\dagger}]_{1},a_{n}]_{2}\hbar^{3}$ produces only
higher-order terms.
The commutation relations $[[a_{0},a_{m}^{\dagger}a_{m}]_{2},a_{n}]_{1}$  contribute  to $[a_{0},a_{n}]_{3}$ in the form $(a_{0}aaa\hbar^{3})$: 
\begin{eqnarray}
 &&\sum_{m=1}{1 \over m}[[a_{0},a_{m}a_{m}^{\dagger}]_{2},a_{n}]_{1}\hbar^{3} \nonumber \\
&=&-{1 \over 8A^{3}}\sum_{k \ne 0,n}\sum_{l \ne 0,n-k}{1 \over n(n-k)\sqrt{|nkl(n-k-l)|}} \nonumber \\
&&\times a_{0}(a_{l}a_{n-k-l}a_{k}+a_{k}a_{l}a_{n-k-l})\hbar^{3} \nonumber \\
&-&{1 \over 4A^{3}}\sum_{m=1}\sum_{k \ne 0,m,m-n}{1 \over m(m-k)\sqrt{|mkn(m-k-n)|}}\nonumber \\
&&\times a_{0}(a_{m-n-k}^{\dagger}a_{k}^{\dagger}+a_{k}^{\dagger}a_{m-n-k}^{\dagger})a_{m}\hbar^{3} \nonumber \\
&+&{1 \over 2A^{3}}\sum_{m=1(m \ne n)}\sum_{k \ne 0,m}{1 \over m^{2}\sqrt{|n(m-n)k(m-k)|}}a_{0}a_{n-m}a_{k}a_{m-k}\hbar^{3} \nonumber \\
&\ominus & {1 \over 4A^{3}}\sum_{m=1(m \ne n)}\sum_{k \ne 0,m-n}{1 \over m(m-n)\sqrt{|mkn(m-k-n)|}}a_{0}a_{k}^{\dagger}a_{m-n-k}^{\dagger}a_{m}\hbar^{3} \nonumber \\
&\oplus & (a_{0}a_{m}^{\dagger}aa+a_{0}aaa_{m+n}+\cdots)\hbar^{3}.
\label{mmn2}
\end{eqnarray}
The terms in the last two lines become higher order, since  $a_{m}a_{n}^{\dagger}|0\rangle =0(m \ne n)$ and $\langle 0 |a_{m}^{\dagger}=a_{m+n}a_{n}^{\dagger}|0\rangle =0 $, etc.. 

\section*{Appendix C \qquad $a_{0}[[a_{0},a_{n}],a_{0}]$}
In this appendix we calculate $a_{0}[[a_{0},a_{n}],a_{0}]$. Expanding
$[a_{0},a_{n}]$ in $\hbar$, we have 
\begin{equation}
a_{0}[[a_{0},a_{n}],a_{0}] 
=a_{0}[[a_{0},a_{n}]_{1},a_{0}]\hbar+ a_{0}[[a_{0},a_{n}]_{2},a_{0}]\hbar^{2}.
\label{in-expansion}
\end{equation}
We calculate these terms by dividing $[a_{0},a_{n}]_{p}(p=1,2)$ into the diagonal and non-diagonal parts.

First, we calculate the first term in (${\rm \ref{in-expansion}}$). The contribution of the diagonal part of $[a_{0},a_{n}]_{1}$  is calculated as
\begin{eqnarray}
&&a_{0}[[a_{0},a_{n}]_{{\rm d}1},a_{0}]\hbar=a_{0}\Bigl[{1 \over nA}a_{0}a_{n},a_{0}\Bigr]\hbar \nonumber \\
 &=&a_{0}{1 \over n}\Bigl[{1 \over A},a_{0}\Bigr]a_{0}a_{n}\hbar- a_{0}{1 \over nA}a_{0}[a_{0},a_{n}]\hbar \nonumber \\
&=&\oplus (a_{0}aaaa_{0}a\hbar^{2}) \oplus  \Bigl(a_{0}{1 \over A}a_{0}^{2}a\hbar^{2}\Bigr)\oplus　\Bigl(a_{0}{1 \over A}a_{0}aa\hbar^{2}\Bigr) \nonumber \\
&=&\ominus (a_{0}aaaa_{0}a\hbar^{2}) \ominus  {1 \over n^{2}A^{2}}a_{0}^{3}a_{n}\hbar^{2} \nonumber \\
&& \ominus {1 \over 2nA^{2}}\sum_{k \ne 0,n}{\epsilon(n) \over \sqrt{|kn(n-k)|}}a_{0}^{2}a_{k}a_{n-k}\hbar^{2}, \end{eqnarray}
where we have used $[{1 \over A},a_{0}] \sim aaa\hbar$. The three terms in the last line only produce higher-order terms. Thus within our approximation the contribution of the diagonal part of $[a_{0},a_{n}]$ is negligible.　The contribution of the non-diagonal part of $[a_{0},a_{n}]_{1}$ is calculated as
\begin{eqnarray}
&&a_{0}[[a_{0},a_{n}]_{{\rm nd}1},a_{0}]\hbar \nonumber \\
&=&a_{0}\sum_{k \ne 0,n}{\epsilon(n) \over 2\sqrt{|kn(n-k)|}}\Bigl[{1 \over A}a_{k}a_{n-k},a_{0}\Bigr]\hbar \nonumber \\
&=&a_{0}\sum_{k \ne 0,n}{\epsilon(n) \over 2\sqrt{|kn(n-k)|}}\Bigl(\Bigl[{1 \over A},a_{0}\Bigr]a_{k}a_{n-k}-{1 \over A}a_{k}[a_{0},a_{n-k}]\nonumber \\
&&-{1 \over A}[a_{0},a_{k}]a_{n-k}\Bigr)\hbar. 
\label{00nnd0}
\end{eqnarray}
The first term in the parentheses in (${\rm \ref{00nnd0}}$) is calculated as
\begin{eqnarray}
&&a_{0}\sum_{k \ne 0,n}{\epsilon(n) \over 2\sqrt{|kn(n-k)|}}\Bigl[{1 \over A},a_{0}\Bigr]a_{k}a_{n-k}\hbar \nonumber \\
&=&\sum_{k \ne 0,n}{\epsilon(n) \over 4A^{2}\sqrt{|kn(n-k)|}}
\sum_{m=1}\sum_{l \ne 0,m}{1 \over m\sqrt{|lm(m-l)|}} 
a_{0}(a_{m}^{\dagger}a_{l}a_{m-l} \nonumber \\
&&-a_{m}a_{l}^{\dagger}a_{m-l}^{\dagger})a_{k}a_{n-k}\hbar^{2} 　
\oplus (a_{0}aaaaa\hbar^{3}).
\label{00n01}
\end{eqnarray}
At the lowest level the second term in the parentheses in (${\rm \ref{00nnd0}}$) is calculated as
\begin{eqnarray}
&-&a_{0}{1 \over 2A}\sum_{k \ne 0,n}{\epsilon(n) \over \sqrt{|kn(n-k)|}}a_{k}[a_{0},a_{n-k}]_{1}\hbar^{2} \nonumber \\
&=& -a_{0}{1 \over 2A^{2}}\sum_{k \ne 0,n}{\epsilon(n) \over (n-k)\sqrt{|kn(n-k)|}}a_{k}{1 \over A}a_{0}a_{n-k}\hbar^{2}  \nonumber \\
& &-a_{0}{1 \over 2A}\sum_{k \ne 0,n}{\epsilon(n) \over \sqrt{|kn(n-k)|}}　\nonumber \\
&& \times a_{k}\sum_{l \ne 0,n-k}{1 \over 2A}{\epsilon(n-k) \over \sqrt{|l(n-k)(n-k-l)|}}a_{l}a_{n-k-l}\hbar^{2} \nonumber \\
&=&-{1 \over 2A^{2}}\sum_{k \ne 0,n}{\epsilon(n) \over (n-k)\sqrt{|kn(n-k)|}}a_{0}^{2}a_{k}a_{n-k}\hbar^{2}　\nonumber \\
&-& {1 \over 4A^{2}}\sum_{k \ne 0,n}\sum_{l \ne 0,n-k}{\epsilon(n) \over (n-k)\sqrt{|knl(n-k-l)|}}a_{0}a_{k}a_{l}a_{n-k-l}\hbar^{2}　\nonumber \\
&+&{1 \over 4A^{3}}\sum_{k \ne 0,n}\sum_{l \ne 0,n-k}{\epsilon(n) \over (n-k)k\sqrt{|knl(n-k-l)|}} \nonumber \\
&&\times a_{0}(a_{k}a_{l}a_{n-k-l}+a_{l}a_{n-k-l}a_{k})\hbar^{3}, 
\label{[0,n-k]1}
\end{eqnarray}
where we have used $\epsilon(k)/|k|=1/k$ and changes of variables $k \rightarrow n-k$.
In (${\rm \ref{[0,n-k]1}}$) we have obtained the two $O(\hbar^{3})$ terms by commuting $a_{k}$ with $a_{0}$ and $1/A$.  At the next level the second term in (${\rm \ref{00nnd0}}$) is calculated from $[a_{0},a_{n-k}]_{2}$ as  
\begin{eqnarray}
&-&a_{0}{1 \over 2A}\sum_{k \ne 0,n}{\epsilon(n) \over \sqrt{|kn(n-k)|}}a_{k}[a_{0},a_{n-k}]_{2}\hbar^{3} \nonumber \\
&=& {1 \over 8A^{3}}\sum_{k \ne 0,n}\sum_{l \ne 0,n-k}{\epsilon(n) \over (n-k)\sqrt{|knl(n-k-l)|}}\Bigl({1 \over l}+{1 \over n-k-l}+{1 \over n-k}\Bigr)\nonumber \\
&& \times a_{0}a_{k}a_{l}a_{n-k-l}\hbar^{3}. 
\label{[0,nk]2}
\end{eqnarray}
The third term in the parentheses in (${\rm \ref{00nnd0}}$) is calculated more easily as
\begin{eqnarray}
&-&a_{0}{1 \over 2A}\sum_{k \ne 0,n}{\epsilon(n) \over \sqrt{|kn(n-k)|}}([a_{0},a_{k}]_{1}a_{n-k}\hbar^{2}+ [a_{0},a_{k}]_{2}a_{n-k}\hbar^{3}) \nonumber \\
&=&-{1 \over 2A}\sum_{k \ne 0,n}{\epsilon(n) \over k\sqrt{|kn(n-k)|}}a_{0}^{2}a_{k}a_{n-k}\hbar^{2} \nonumber \\　
&-& {1 \over 4A^{2}}\sum_{k \ne 0,n}\sum_{l \ne 0,n-k}{\epsilon(n) \over (n-k)\sqrt{|knl(n-k-l)|}}a_{0}a_{l}a_{n-k-l}a_{k}\hbar^{2} \nonumber \\
&+& {1 \over 8A^{3}}\sum_{k \ne 0,n}\sum_{l \ne 0,n-k}{\epsilon(n) \over (n-k)\sqrt{|knl(n-k-l)|}}\Bigl({1 \over l}+{1 \over n-k-l}+{1 \over n-k}\Bigr) \nonumber \\ 
&& \times a_{0}a_{l}a_{n-k-l}a_{k}\hbar^{3},
\label{[0,k]1}
\end{eqnarray}
where we have used changes of variables $k \rightarrow n-k$.

Second, we calculate the second term in (${\rm \ref{in-expansion}}$).
As for the lowest term \\
$a_{0}[[a_{0},a_{n}]_{2},a_{0}]_{1}\hbar^{3}$, only the non-diagonal interactions give non-zero contributions
\begin{eqnarray}
&&a_{0}[ [a_{0},a_{n}]_{{\rm nd}2},a_{0}]_{{\rm nd}1}\hbar^{3} \nonumber \\
&=& a_{0}\Bigr[a_{0},{\epsilon(n) \over 4A^{2}}\sum_{k \ne 0,n}\Bigl({1 \over k}+{1 \over n-k}+{1 \over n}\Bigr){1 \over \sqrt{|nk(n-k)|}}a_{k}a_{n-k}\Bigr]_{{\rm nd}1}\hbar^{3} \nonumber \\
&=&{\epsilon(n) \over 8A^{3}}\sum_{k \ne 0,n}\sum_{l \ne 0,k}{1 \over (n-k)\sqrt{|nkl(n-k-l)|}}\Bigl({1 \over n}+{1 \over k}+{1 \over n-k}\Bigr) \nonumber \\
&\times&a_{0}(a_{k}a_{l}a_{n-k-l}
+a_{l}a_{n-k-l}a_{k})\hbar^{3}.
\label{a021} 
\end{eqnarray}
Other terms in the third term in (${\rm \ref{in-expansion}}$) give higher-order terms.

Collecting (${\rm \ref{00n01}}$)$\sim$(${\rm \ref{a021}}$), we have
\begin{eqnarray}
&&a_{0}[[a_{0},a_{n}],a_{0}] \nonumber \\
&=&\sum_{k \ne 0,n}\sum_{m=1}\sum_{l \ne 0,m}
{\epsilon(n) \over 4A^{2}m\sqrt{|kn(n-k)lm(m-l)|}} \nonumber \\
&\times&a_{0}(a_{m}^{\dagger}a_{l}a_{m-l}-a_{m}a_{l}^{\dagger}a_{m-l}^{\dagger})a_{k}a_{n-k}\hbar^{2} \nonumber \\
&-&{1 \over 4A^{2}}\sum_{k \ne 0,n}\sum_{l \ne 0,n-k}{\epsilon(n) \over (n-k)\sqrt{|knl(n-k-l)|}} \nonumber \\
&& \times a_{0}(a_{k}a_{l}a_{n-k-l}+a_{l}a_{n-k-l}a_{k})\hbar^{2}　\nonumber \\
&-&{\epsilon(n) \over 2A^{2}}\sum_{k \ne 0,n}\Bigl({1 \over n}+{1 \over k}+{1 \over n-k}\Bigr){1 \over \sqrt{|kn(n-k)|}}a_{0}^{2}a_{k}a_{n-k}\hbar^{2}　\nonumber \\
&+&{\epsilon(n) \over 8A^{3}}\sum_{k \ne 0,n}\sum_{l \ne 0,n-k}{1 \over (n-k)\sqrt{|nkl(n-k-l)|}}\Bigl({1 \over n}+{3 \over k}+{2 \over n-k}+{1 \over l} \nonumber \\
&&+{1 \over n-k-l}\Bigr)a_{0}(a_{k}a_{l}a_{n-k-l}+a_{l}a_{n-k-l}a_{k})\hbar^{3}. 
\label{00n0}
\end{eqnarray}
\section*{Appendix D \qquad $[[[a_{0},a_{n}],a_{0}],a_{0}]$}
The lowest-order term in $[[[a_{0},a_{n}],a_{0}],a_{0}]$ is $[[[a_{0},a_{n}]_{1},a_{0}]_{1},a_{0}]_{1}\hbar^{3}$. 
In these commutation
relations each $[{1 \over A},a_{0}]_{1}$ produces only higher-order terms.
There are only three ways to get relevant terms.  Choosing one diagonal interaction and two non-diagonal interactions for the three brackets, we have
\begin{eqnarray}
&&[[[a_{0},a_{n}]_{1},a_{0}]_{1},a_{0}]_{1}\hbar^{3} \nonumber \\
=&&{1 \over 4A^{3}}\sum_{k \ne 0,n}\sum_{l \ne 0,n-k}{\epsilon(n)\epsilon(n-k)
\over |n-k|\sqrt{|nkl(n-k-l)|}} \nonumber \\
&\times &\Bigl( \bigl({1 \over n}+{1 \over l}+{1 \over n-k-l}+{1 \over k}\bigr)a_{0}(a_{l}a_{n-k-l}a_{k}+a_{k}a_{l}a_{n-k-l})\nonumber \\
&+& \bigl({2 \over n-k}a_{0}a_{l}a_{n-k-l}a_{k}+{2 \over k}a_{0}a_{k}a_{l}a_{n-k-l}\bigr)\Bigr)\hbar^{3}. 
\label{n000}
\end{eqnarray}
The other terms in $[[[a_{0},a_{n}],a_{0}],a_{0}]$   produce only higher-order terms.

\section*{Appendix E \qquad $[[[a_{0},a_{n}],a_{m}],a_{m}^{\dagger}]$}
First, we calculate the lowest-order term in $[[[a_{0},a_{n}],a_{m}],a_{m}^{\dagger}]$:
\begin{eqnarray}
&&[[[a_{0},a_{n}]_{1},a_{m}]_{1},a_{m}^{\dagger}]_{1}\hbar^{3} 
=\Bigl[\Bigl[{1 \over nA}a_{0}a_{n},a_{m}\Bigr]_{1},a_{m}^{\dagger}\Bigr]_{1}\hbar^{3} \nonumber \\
&&+\sum_{k \ne 0,n}{\epsilon(n) \over \sqrt{|kn(n-k)|}}\Bigl[\Bigl[{1 \over 2A}a_{k}a_{n-k},a_{m}\Bigr]_{1},a_{m}^{\dagger}\Bigr]_{1}\hbar^{3}. 
\label{d-and-n3}
\end{eqnarray}
In the first term in (${\rm \ref{d-and-n3}}$) the non-diagonal interactions give
higher-order terms. From the diagonal interactions we have
\begin{eqnarray}
&&\Bigl[\Bigl[{1 \over nA}a_{0}a_{n},a_{m}\Bigr]_{1},a_{m}^{\dagger}\Bigr]_{1}\hbar^{3} \nonumber \\
&=&{2 \over nmA^{3}}a_{0}a_{n}\hbar^{3}+\delta_{n,m}{1 \over n^{2}A^{2}}(a_{n}a_{0}+a_{0}a_{n})\hbar^{3} 
-{1 \over nm^{2}A^{3}}a_{0}a_{m}a_{m}^{\dagger}a_{n}\hbar^{3} \nonumber \\
&&\oplus  (aa+a_{0}aaa_{n+m})\hbar^{3} \nonumber \\
&=&{2 \over nmA^{3}}a_{0}a_{n}\hbar^{3}+\delta_{n,m}{2 \over n^{2}A^{2}}a_{0}a_{n}\hbar^{3}-\delta_{n,m}{1 \over n^{3}A^{3}}a_{0}a_{n}\hbar^{4}  \nonumber \\
&&-{1 \over nm^{2}A^{3}}a_{0}a_{n}\hbar^{4}\oplus (aa+a_{0}aaa_{n+m}+a_{0}a_{m}^{\dagger}aa)\hbar^{3}, n>0.
\label{diag3}
\end{eqnarray}
As for the second term in (${\rm \ref{d-and-n3}}$) only the non-diagonal  interactions in $[{1 \over A},a_{m}]$  produce non-zero contributions
\begin{eqnarray}
&&\sum_{k \ne 0,n}{\epsilon(n) \over \sqrt{|kn(n-k)|}}\Bigl[\Bigl[{1 \over 2A}a_{k}a_{n-k},a_{m}\Bigr]_{1},a_{m}^{\dagger}\Bigr]_{1}\hbar^{3} \nonumber \\
&=&-{1 \over 2A^{3}}\sum_{l \ne 0,m}(1-\delta_{n,m}){\epsilon(n)\epsilon(m) \over
m\sqrt{|ln(m-l)(n-m)|}}a_{0}a_{l}a_{m-l}a_{n-m}\hbar^{3}, \nonumber \\
&& \qquad \qquad \qquad \qquad \qquad \qquad \qquad \qquad n>0. \label{nond3}
\end{eqnarray}

Second, we calculate the next-order term $[[[a_{0},a_{n}]_{i},a_{m}]_{j},a_{m}^{\dagger}]_{k}\hbar^{4}\quad (i+j+k=4)$. In these commutation relations only
the diagonal interactions produce non-zero contributions. Paying attention
to the commutation relations $[{1 \over A},a_{m}]$, we have
\begin{eqnarray}
&&[[[a_{0},a_{n}]_{2},a_{m}]_{1},a_{m}^{\dagger}]_{1} +[[[a_{0},a_{n}]_{1},a_{m}]_{2},a_{m}^{\dagger}]_{1} \nonumber \\
&=&-\Bigl[\Bigl[{1 \over n^{2}A^{2}}a_{0}a_{n},a_{m}\Bigr]_{1},a_{m}^{\dagger}\Bigr]_{1}
+ \Bigl[\Bigl[{1 \over nA}a_{0}a_{n},a_{m}\Bigr]_{2},a_{m}^{\dagger}\Bigr]_{1}
\nonumber \\
&=&-\Bigl({3 \over n^{2}m}+{2 \over nm^{2}}\Bigr)(1+\delta_{n,m}){1 \over A^{3}} a_{0}a_{m}\hbar^{4}, n>0.
\label{0nmm4}
\end{eqnarray}
The rest of $O(\hbar^{4})$ terms $[[[a_{0},a_{n}]_{1},a_{m}]_{1},a_{m}^{\dagger}]_{2}\hbar^{4}$ produces
only higher-order contributions like $(a_{0}aaa\hbar^{4})$.
From (${\rm \ref{diag3}}$),(${\rm \ref{nond3}}$) and  (${\rm \ref{0nmm4}}$) we have
\begin{eqnarray}
&&{1 \over 6A}\sum_{m=1}{1 \over m}[[[a_{0},a_{n}],a_{m}],a_{m}^{\dagger}] \nonumber \\
&=&{\zeta(2) \over 3nA^{3}}a_{0}a_{n}\hbar^{3}+{1 \over 3n^{3}A^{3}}a_{0}a_{n}\hbar^{3} \nonumber \\
&-& \sum_{m=1,m\ne n}\sum_{l \ne 0,m}{1 \over 12m^{2}\sqrt{|ln(m-l)(n-m)|}A^{4}}a_{0}a_{l}a_{m-l}a_{n-m}\hbar^{3} \nonumber \\
&-&{1 \over n^{4}A^{4}}a_{0}a_{n}\hbar^{4} 
-{\zeta(2) \over 2n^{2}A^{4}}a_{0}a_{n}\hbar^{4}-{\zeta(3) \over 2nA^{4}}a_{0}a_{n}\hbar^{4}, n>0.
\label{0nmm}
\end{eqnarray}
 \\
{References}\\
1)For a review, see S.~J.~Brodsky, H-C.~Pauli and S.~S.~Pinsky, Phys.~Rep. {\bf 301} (1998), 299. \\
2)S.~J.~Chang, R.~G.~Root and T.~M.~Yan, Phys.~Rev.~D7,1973,1133. \\  
        S.~J.~Chang and T.~M.~Yan,  Phys.~Rev.~{D7,1973,1147}.\\
3)T.~Maskawa and K.~Yamawaki, Prog.~Thor.~Phys.~{56,1976,270}.\\
4)S.~J.~Chang, Phys.~Rev.~{D13,1976,2778}.\\
J.~Abad, J.~G.~Esteve and A.~F.~Pacheco, Phys.~Rev.~ {D32,1985,2729}.\\
M.~Funke, U.~Kaulfuss and H.~Kummel, Phys.~Rev.~ {D35,1987,621}.\\
H.~Kroger, R.~Girard and G.~Dufour,  Phys.~Rev.~{D35,1987,3944}.\\
5)T.~Heinzl, S.~Krusche, S.~Simburger and E.~Werner, Z.~Phys.{\bf C56}(1992), 415.\\ 
  T.~Heinzl, S.~Krusche and E.~Werner, Phys.~Lett.~{B272,1991,54}.\\
6)D.~G.~Robertson, Phys.~Rev.~{D47,1993,2549}.\\
7)C.~M.~Bender, S.~S.~Pinsky and B.~van de Sande, Phys.~Rev.~{D48,1993,816}.
   S.~S.~Pinsky and B.~van de Sande, Phys.~Rev.~{D49,1994,2001}.\\
8)S.~Pinsky, B.~van de Sande and J.~R.~Hiller,  Phys.~Rev.~{D51,1995,726}.
9)K.~Oshima and M.~Yahiro, Prog.~Thor.~Phys.~{101,1999,459}.\\
10) To discretize the momentum space is advantageous also in calculating
the hadron spectra; see, H.~-C.~Pauli and S~.J.~Brodsky, {D32,1985,1993, 2001}.\\
11) C.~M.~Bender, L~.R.~Mead and S.~S.~Pinsky, Phys.~Rev.~Lett.~{56,1986,2445}.
12)J.~Killingbeck, {\it Microcomputer Quantum Mechanics}(Adam Hilger, Bristol, 1983).\\
\end{document}